\documentclass[12pt]{article}

\def\px{\hat{p}_x}
\def\x{\hat{x}}
\def\peta{\hat{p}_\eta}

\def\la{\langle}
\def\ra{\rangle}
\def\et{\hat{\eta}}
\def\a{\hat{a}}
\def\b{\hat{b}}
\def\xpb{\hat{x}_{pb}}
\def\ppb{\hat{p}_{pb}}
\def\xs{\hat{x}_s}
\def\pxs{\hat{p}_s}

\def\w{\hat{w}}
\def\xa{x_{s}}
\def\ya{p_{s}}
\def\xb{x_{pb}}
\def\yb{p_{pb}}
\usepackage{amssymb}
\usepackage{graphicx}
\usepackage{rotating}
\usepackage{times}

\usepackage{scicite}

\usepackage{times}

\newenvironment{sciabstract}{%
\begin{quote} \bf}
{\end{quote}}

\newcounter{lastnote}
\newenvironment{scilastnote}{%
\setcounter{lastnote}{\value{enumiv}}%
\addtocounter{lastnote}{+1}%
\begin{list}%
{\arabic{lastnote}.}
{\setlength{\leftmargin}{.22in}}
{\setlength{\labelsep}{.5em}}}
{\end{list}}

\title{The Quantum Mellin transform}

\author
{J. Twamley$^{1\ast}$ and G. J. Milburn$^2$\\
\\
\normalsize{$^{1}$Centre for Quantum Computer Technology, Macquarie University}\\ \normalsize{Sydney, New South Wales 2109, Australia,}\\
\normalsize{$^{2}$Centre for Quantum Computer Technology, The University
of Queensland}\\\normalsize{St Lucia, QLD 4072, Australia}\\
\\
\normalsize{$^\ast$To whom correspondence should be addressed; E-mail:  jason.twamley@mq.edu.au}
}

\date{}

\begin{document} 




\maketitle


\begin{sciabstract}
  We uncover a new type of unitary operation for quantum mechanics on the half-line which yields a transformation to ``Hyperbolic phase space'' $(\eta, p_\eta)$. 
We show that this new unitary change of basis from the position $x$ on the half line to the Hyperbolic momentum $p_\eta$, transforms the wavefunction via a Mellin transform on to the critial line $s=1/2-ip_\eta$. We utilise this new transform to find quantum wavefunctions whose Hyperbolic momentum representation approximate a class of higher transcendental functions, and in particular, approximate the Riemann Zeta function. We finally give possible physical realisations to perform an indirect measurement of the Hyperbolic momentum of a quantum system on the half-line.  
\end{sciabstract}


\section{Introduction} Higher transcendental functions find many uses in mathematics, engineering, physics and many other sciences. Their efficient numerical evaluation is a highly challenging task and typically one makes
use of pre-compiled library routines \cite{thompson97}. In this work we present a curious method which can {\em design} a 1D quantum system in such a way that the resulting wavefunction is proportional to a given transcendental function taken from a certain class of functions. The resulting quantum system displays enormously rich behaviour. By studying the corresponding quantum system one might discover new insights into the properties these transcendental functions. 
More interestingly we discover how to execute a new quantum unitary transform, the quantum Mellin transform, (similar to the quantum Fourier transform), which takes the Mellin transform of a wavefunction. The quantum Fourier transform has been prominent in many quantum algorithms and one might suspect that the quantum Mellin transform might also be useful in quantum computation.

The class of transcendental functions we consider are the three parameter family of higher transcendental functions sometimes known as the
 \textit{Lerch transcendents} $\Phi(z,s,u)$, which contain the celebrated Riemann Zeta function $\zeta(s)$ ( \cite{bateman53} section 1.11 and \cite{spanier87} section 64:12). The Lerch transcendent is given as
the analytic continuation of the series
\begin{equation}\
\Phi(z,s,u)=\sum_{n=0}^\infty\, \frac{z^n}{(u+n)^s}\;\;,|z|<1,\;\;u\ne -1, -2, -3, \cdots
\end{equation}
which converges for $u\in\mathbb{R}^+$, $z,s\in \mathbb{C}$, with either $(|z|<1,\Re(s)>0)$,  or $(|z|=1,
\, \Re(s)>1)$.
Special cases include the analytic continuations of the
\textit{Riemann zeta function} and the \textit{Hurwitz zeta function}, (valid for $\Re(s)>1$),
\begin{equation}\label{fzeta}
\zeta(s)=\sum_{k=1}^{\infty} {1 \over k^s}=\Phi(1,s,1),\;\;
\zeta(s,u)=\sum_{k=0}^{\infty} {1 \over (u+k)^s}=\Phi(1,s,u),
\end{equation}
the \textit{alternating zeta function} (also known as
\textit{Dirichlet's eta function}), and the the \textit{Dirichlet beta function}, (valid for $Re(s)>0$),
\begin{eqnarray}\label{alter}
\zeta^{*}(s)&&= \sum_{k=1}^{\infty} {(-1)^{k-1} \over k^s} =
\Phi(-1,s,1),\\
\beta(s)&&=\sum_{k=0}^{\infty} {(-1)^k \over (2k+1)^s}= 2^{-s} \Phi
\left(-1,s,{1 \over 2} \right),
\end{eqnarray}
the \textit{Polylogarithm}, and the \textit{Lerch zeta function}
\begin{equation}
{\textrm{Li}_n(z)}=\sum_{k=1}^{\infty} {z^k \over k^n}=z
\Phi(z,n,1),\;\;
L(\lambda,\alpha,s)=\Phi(\exp(2 \pi i \lambda),s,\alpha).
\end{equation}

In the following we will be particularly interested in the evaluation of Lerch transcendent on the  {\it critical line}, when $s=1/2+it$, and for this case $|z|<1$. Thus the special cases of (\ref{fzeta}), do not converge while those of (\ref{alter}), do converge. It is known that the Lerch transcendent can be expressed as a Mellin transform: if  ($|z|<1$, $\Re(s)>0$), or ($z=1$, $\Re(s)>1$), then
\begin{equation} \label{mellinphi}
\Delta(z,s,u)\equiv\Gamma(s)\Phi(z,s,u)= \int_{0}^{\infty} {e^{-(u-1)t}
\over e^t-z} t^{s-1} dt, \label{mellinlerch}
\end{equation}
and making use of the relation $\zeta^{*}(s)=(1-2^{1-s})\zeta(s)$, we have the very interesting special case, where the Riemann Zeta function can be expressed as a Mellin transform,
\begin{equation}
(1-2^{1-s})\zeta(s)=\zeta^*(s)=\Phi(-1,s,1)=\frac{1}{\Gamma(s)}\int_{0}^{\infty} 
\frac{1} {e^t+1} t^{s-1} dt,\label{alterzeta} 
\end{equation}

\par In summary, we proceed below to uncover a new unitary change of basis in the case of quantum mechanics on $\mathbb{R}^+$, which corresponds to the evaluation of the Mellin transform of a wavefunction when $s=1/2+it$. This transform corresponds to moving from the $|x\rangle$, $(x\in\mathbb{R}^+$), basis to the $|p_\eta\rangle$, ($p_\eta\in\mathbb{R}$), basis where we call the latter the Hyperbolic momentum basis. We then deduce suitable quantum mechanical potentials and boundary conditions that yield ground state wavefunctions which, when viewed in the Hyperbolic momentum basis, give the left hand side of  (\ref{mellinphi}), when $s=1/2+it$ which includes the Riemann-Zeta function as a special case. For the Riemann-Zeta function case we find that the relevant wavefunction $\psi_\zeta(x),\;x\in\mathbb{R}^+$, is quite simple and corresponds to the ground (bound) state of a deceptively simple 1D potential. Such a potential might be physically engineered in a variety of systems, e.g. neutral atom mirror traps, trapped ion systems.

As a second case we can also find {\em unbounded}, wavefunctions in the $|x\rangle$  basis which, when viewed in the $|p_\eta\rangle$ basis, give the Riemann Zeta function $\zeta(1/2-ip_\eta)$, precisely. In this case both the original and transformed  states are unbounded. However as the basis transform is unitary, and thus preserves inner products, we would conjecture that the analytic continuation of these states gives results identical to the analytic continuation of the Riemann Zeta function onto the critical line. 

Finally we show how one can obtain bounded wavefunctions which can approximate the Riemann Zeta function in a more systematic (though somewhat complicated) fashion. We finally present a scheme based on the Degenerate Parametric Oscillator to measure the Hyperbolic momentum in a physical system corresponding to the Degenerate Parametric Oscillator. 


One  interesting result of this work is that we are able to find a physical quantum system whose ground state wavefunction, when viewed in the appropriate basis, possesses zeros whose locations exactly match those of the Riemann Zeta function on the critical line.  This tantalising result can be
compared with the Hilbert-Polya formulation of the Riemann Hypothesis: to find a self-adjoint linear Hermitian operator whose spectrum matches the locations of the zeros of the Riemann Zeta function on the critical line \cite{montgomery74}.  Evidence for the latter seems quite strong following extensive numerical analysis of the spacings between adjacent zeros of the Riemann Zeta function and their comparison with the statistics of a Generalized Unitary Ensemble (GUE - or eigenvalues of random Hermitian matrices) \cite{montgomery74, odlyzko89,rudnick96,katz99, NoteAmerMath03}. It is not clear to us how our construction has relevance to the Hilbert-Polya formulation of the Riemann-Zeta hypothesis but the existence of relatively ``simple'' quantum wavefunctions with randomly distributed zeros seems intriguing. 

\section{Quantum Mechanics on $\mathbb{R}^+$}
\label{QMinhalfline}
We now turn to the description of quantum mechanics on the half-line $x\in \mathbb{R}^+$. There have been numerous expositions of the quantum mechanics of a particle moving in $\mathbb{R}^+$. Almost without exception, however, these analyses are executed in the position basis $|x\rangle$.  Here, we instead seek to make a closer correspondence between the conjugate phase-space operators $\hat{x},\,\hat{p}$, $[\hat{x},\hat{p}]=i\hbar$, for a particle moving on the full-line  $x\in\mathbb{R}$, with analogous conjugate  phase-space operators on the half-line. As the latter must represent physical observables they must be represented by Hermitian Self-Adjoint operators.  To determine whether an operator $\hat{A}$, is Hermitian one must examine the action of the operator  $\hat{A}$, and its adjoint $\hat{A}^\dagger$, but also the $L^2$ domains ($\psi(x) \in L^2[0,\infty]$, if $\int_0^\infty|\psi(x)|^2dx$ is finite), of these operators, $D(\hat{A})$, and $D(\hat{A}^\dagger)$. If the actions of $\hat{A}$ and $\hat{A}^\dagger$ are identical, i.e. $(\hat{A}^\dagger\phi,\psi)=(\phi,\hat{A}\psi),\;
\forall \phi,\psi\in D(\hat{A})$, then the operator $\hat{A}$, is Hermitian provided that $D(\hat{A})\in D(\hat{A}^\dagger)$. Further it is Self-Adjoint if the actions match and so too do the domains of the operator and its adjoint, $D(\hat{A}^\dagger)=D(\hat{A})$. An operator can be Hermitian but not Self-Adjoint and this will give rise to unphysical situations in that the action of $\hat{A}^\dagger$ may change the domain of the wavefunction, i.e. to different boundary conditions etc.  

Here we follow closely the work of \cite{twamley98}, and \cite{milburn94}, to find that the obvious candidate for the momentum operator $\hat{p}_x\sim -i\hbar d/dx$, conjugate to $x\in\mathbb{R}^+$,  is not Hermitian and Self-Adjoint. Moreover, there is no self-adjoint extension possible for this operator. This is to be contrasted to the case of a particle confined to a finite interval $x\in[a,b]$ \cite{carreau90,daluz94},  or on the entire real line with delta-function potentials or with the origin removed \cite{araujo04,bonneau01}. We instead proceed to define a new momentum operator $\hat{p}_\eta$, the ``Hyperbolic momentum'' which corresponds to dilations of the $\hat{x}$ coordinate or displacements of the ``Hyperbolic position'' $\eta\equiv \ln x$, coordinate.

Before examining the case of $\hat{p}_x$, let us examine the somewhat easier case of the kinetic energy operator on the half-line,
\begin{equation}
H_0=-\frac{\hbar^2}{2m}\frac{d^2}{dx^2}\;\;.
\end{equation}
As mentioned above we have to pay particular attention to the domain of operators on which $H_0$ is allowed to act. We define $\psi(x)\in \Omega\equiv C_2^\infty([0,\infty])$, to be the class of functions which are infinitely differentiable (absolutely continuous, i.e. $f(x)=\int_0^x(df/dx)dx+f(0)$), and have finite $L^2$ norm on the interval $[0,\infty]$. 
If we now consider two functions $\phi_1, \phi_2\in\Omega$, we can examine the Hemiticity of $H_0$, by integrating by parts,
\begin{eqnarray}
(H_0^\dagger\phi_2,\phi_1)-(\phi_2,H_0\phi_1)&=&\phi_2^*(0)\left.\frac{d\phi_1}{dx}\right|_{x=0}-
\left.\frac{d\phi_2^*}{dx}\right|_{x=0}\phi_1(0)\;\;,\\
&=&\phi^*_2(0)\phi_1(0)\left[\frac{\phi_1^\prime}{\phi_1}-\frac{\phi_2^\prime}{\phi_2}\right]_{x=0}\;\;.\label{herm1}
\end{eqnarray}
If one chooses boundary conditions (A) $\mathbb{C_A}\equiv\phi_2^*(0)=\phi_1(0)=0$, or (B) $\mathbb{C}_B\equiv \phi_1(0)=d\phi_1^*/dx|_0=0$, then $H_0$ is Hermitian. However in case (A) the domains of $H_0$ and $H_0^\dagger$ are fixed to be identical by $\mathbb{C}_A$, and thus $H_0$ is also self-adjoint. In case (B)   $D(H_0)$ is reduced to  $\phi_1\in\Omega \cap \mathbb{C}_B $,  while $D(H_0^\dagger)=\Omega$, and thus the operator $H_0$ is not self-adjoint. From (\ref{herm1}), a more general class of boundary conditions which give an Hermitian and self-adjoint $H_0$, is when $\frac{\phi_1^\prime(0)}{\phi_1(0)}=\kappa\in\mathbb{R}$.  This choice of boundary condition extends the operator $H_0$, into a one-parameter class of operators and this process is known as a self-adjoint extension.

If we now repeat the above analysis for the operator $\hat{p}_x=-i\hbar d/dx$, again assuming $\phi_1,\phi_2 \in \Omega$, we have
\begin{eqnarray}
(\hat{p}_x^\dagger \phi_2,\phi_1)-(\phi_2,\hat{p}_x\phi_1)&=&i\hbar\int_0^\infty (\phi_2^{*\,\prime}\phi_1+\phi_2^*\phi_1^{\prime})dx\;\;,\\
&=&i\hbar\left[\left.\phi_2^*\phi_1\right|_\infty-\left.\phi_2^*\phi_1\right|_0\right]\;\;.\label{P1}
\end{eqnarray}
The first term in (\ref{P1}) vanishes due to the $L^2$ nature of $\Omega$, and thus for the second term to vanish one must set either $\phi_1(0)=0$, {\em or} $\phi_2(0)=0$. In either case the resulting operator $\hat{p}_x$, will be Hermitian but will not be self-adjoint. It is not obvious that we can choose alternative boundary conditions that will yield a self-adjoint extension of this non-self-adjoint operator. Fortunately there is a theorem of von Neumann \cite{vonneumann29,bonneau01}, which tells us whether such an extension exists or not.  The method is quite simple and involves comparing the dimensions of the so-called, ``deficiency subspaces'', ${\cal N}_\pm$, of the operator in question, e.g. $\hat{A}$. These spaces are defined by
\begin{equation}
{\cal N}_\pm=\left\{\psi\in D(\hat{A}^\dagger),\;\;\hat{A}^\dagger\psi=z_\pm \psi,\;\;\Im(z_\pm) \gtrless\, 0\right\}\;\;,
\end{equation}
and to find $(n_+,n_-)$, the dimensions of these spaces, one just looks for the number of normalisable independent solutions to $\hat{A}^\dagger \psi=\pm i\gamma \psi$, for $\gamma$ real and positive. For $\hat{p}_x=-i\hbar d/dx$, we easily have $\psi_\pm\sim \exp(\pm\gamma x/\hbar)$, and for $x\in\mathbb{R}^+$, $(n_+,n_-)=(0,1)$. From \cite{vonneumann29}, if the deficiency indices are different then the operator is not self-adjoint and moreover {\em no self-adjoint extension is possible}. If we instead had considered the case of the whole real line then $(n_+,n_-)=(0,0)$, i.e. neither wavefunction is normalisable on the entire real line, and in this case not only is $\hat{p}_x$, self-adjoint, it is termed {\em essentially self-adjoint} as its domain is the entire $\phi\in L^2[-\infty,+\infty]$. 

\section{Hyperbolic Coordinates: Dilatons $\rightarrow$ Displacements}
That the typical expression for the momentum operator $\hat{p}_x=-i\hbar d/dx$, does {\em not} correspond to a physical operator on $\mathbb{R}^+$, seems to have only generated little attention in the literature. On the other hand, as $[\x,\px]=i\hbar$, the action of $\hat{p}_x$ on $\hat{x}$, yields a displacement, 
\begin{equation}
e^{i\alpha \px/\hbar}\x e^{-i\alpha\px/\hbar}=\x+\alpha\;\;,
\end{equation}
and this  cannot represent a physical operation as it alters the domain $x\in\mathbb{R}^+$. Instead we examine the operator which generates dilations of the $\x$ operator:
\begin{eqnarray}
\hat{p}_\eta \equiv \frac{1}{2}\left[ \x\px+\px\x\right]&=&\x\px -i\frac{\hbar}{2}\;\;,\\
&=&-i\hbar\left[ x\frac{d}{dx}+\frac{1}{2}\right]\;\;,\label{pdilaton_x}
\end{eqnarray}
where the last line is $\peta$ evaluated in the $\x$ representation.  This operator, as a generator of spatial dilations has already been studied as part of a larger space-time conformal transformation in certain non-relativistic quantum mechanical problems by de Alfaro, Fubini and Furlan \cite{dealfaro76}, and Jackiw \cite{Jackiw91}. The operator also appears in work by van Winter \cite{vanwinter98}, and Berry and Keating  \cite{berry99}, where they take this operator to be the Hamiltonian of a system. Indeed, although connections between the Riemann Zeta function and eigenfunctions of the operator $H=\x\px$ were made in \cite{berry99,berry99a}, our treatment differs significantly in that we consider the operator $p_\eta=\x\px+\px\x$, to be a new conjugate variable describing dilation/conformal momentum. From $[\x,\peta]=i\hbar\x$, the Sack algebra \cite{sack58}, we can see that the action of this new momentum on $\x$, is as expected:
\begin{equation}
e^{i\mu\peta/\hbar}\x e^{-i\mu\peta/\hbar}=\x e^{\mu}\;\;.
\end{equation}
To test for Hermiticity we examine
\begin{equation}
K=(\peta^\dagger\phi_1,\phi_2)-(\phi_1,\peta\phi_2)\;\;,
\end{equation}
and by setting $\phi_1=U/\sqrt{x},\;\phi_2=V/\sqrt{x}$, we find 
\begin{equation}
K=\left. U^*V\right|_0^\infty\;\;,
\end{equation}
which vanishes if $V(0)=U(0)=0$, and thus $\peta$ is Hermitian for $\phi\sim U(x)/\sqrt{x},\; \{\phi\in \bar{\Omega}\equiv L^2[\mathbb{R}^+, dx, U(0)=0]\}$.  To see if the operator is self-adjoint we use the von Neumann test and examine $\peta\psi=\pm i\gamma\psi$. We find $\psi\sim V/\sqrt{x}$, but $V(x)=A_\pm x^{\mp \gamma/\hbar}$, neither of which are in $\bar{\Omega}$, and thus we have the deficiency indices $(n_+,n_-)=(0,0)$, indicating that $\peta$ is an {\em essentially self-adjoint} operator. Defining $\hat{\eta}=\ln \x$, (this will be made more precise later on), we can recover the standard Heisenberg algebra and displacement operation, albeit in the ``exponent space'': $[\et, \peta]=i\hbar$, and
\begin{equation}
e^{i\alpha\peta/\hbar}\et e^{-i\alpha\peta/\hbar}=\et+\alpha\;\;.
\end{equation} 
Thus by making the unitary transformation of the quantum mechanics on the half line described by the ``conjugate operators'', $(\x,\peta)$, to the exponential representation on the full-line, $(\ln \x, \peta)$, we can regain the familiar Heisenberg algebra. Dilations of $\x$ now become displacements of $\eta\equiv \ln \x$, and one can formulate all of the standard QM on $\mathbb{R}^+$, in this Hyperbolic phase-space.

We now evaluate the transition matrix elements $\la p_\eta |x\ra,\,\la p_\eta |\eta\ra,\,\la p_\eta |\psi\ra$, and resolutions of unity in the new basis in order to obtain the correct measures. We will also obtain a displacement operator $D(\lambda,\mu)$, in the $(\eta,\peta)$, phase-space and from this a Wigner function pseudo-probability representation of a wavefunction in this phase-space. Armed with these tools we will then be in a position to seek wavefunctions where $\la p_\eta | \psi\ra$ are related to the Lerch transcendent (and Riemann Zeta function in particular), on the critical line. First we look at the eigenstates of $\peta$ by taking $\la x | \peta |p_\eta\ra=p_\eta\la x |p_\eta\ra$, and where $\psi^{p_\eta}(x)=V^{p_\eta}(x)/\sqrt{x}$, to obtain
\begin{equation}
\la x|p_\eta\ra=\frac{1}{\sqrt{2\pi}}x^{ip_\eta/\hbar-\frac{1}{2}}\;\;.\label{p_eigen1}
\end{equation}
As $[\et,\peta]=i\hbar$, we can define the eigenstates of the coordinate conjugate to $\peta$ via the Fourier transform
\begin{equation}
|\eta\ra=\frac{1}{\sqrt{2\pi\hbar}}\int_{-\infty}^{+\infty}\,dp_\eta\, e^{-ip_\eta\eta/\hbar}|p_\eta\ra\;\;,
\end{equation}
which gives
\begin{eqnarray}
\la x|\eta\ra&=&\frac{1}{2\pi\hbar}\int_{-\infty}^{+\infty}dp_\eta\,e^{-ip_\eta\eta}\frac{x^{ip_\eta/\hbar}}{\sqrt{x}}\;\;,\\
&=&\frac{e^{-u/2}}{2\pi\hbar}\int_{-\infty}^{+\infty}dp_\eta\, e^{-ip_\eta(\eta-u)/\hbar}
=e^{-u/2}\delta(\eta-u)=\sqrt{x}\delta(x-e^\eta)\;\;,\label{x_inner}
\end{eqnarray}
where we have set $u=\ln x$, and have used $\delta(u)/g^\prime(u)=\delta(x), \, x=g(u)$. We see that the $\et$ eigenstates are not merely rescalings of the $\x$ eigenstates but also include a square-root weighting. From (\ref{x_inner}), we see that 
\begin{equation}
\la \eta | \psi\ra=\int_0^\infty dx \,\psi(x) \sqrt{x}\delta(x-e^\eta)=e^{\eta/2} \psi(x=e^{\eta})\;\;.\label{psi_x_inner}
\end{equation}
Let us define the following ket so as to absorb the square-root factor:
\begin{equation}
\overline{|\eta\ra}\equiv \sqrt{e^\eta}|x=e^\eta\ra_x\;\;,
\end{equation}
where we have inserted the subscript $x$ to emphasise that the ket in question is evaluated in the $x$ basis but at the value where $x=\exp(\eta)$. Using this notation, (\ref{p_eigen1}), and the properties of the $\delta-$function we can show
\begin{eqnarray}
\la p_\eta^1|p_\eta^2\ra&=&\int_{-\infty}^{+\infty}d\eta\, \la p_\eta^1|\overline{|\eta\ra}\overline{\la \eta|}|p_\eta^2\ra\;\;,\\
&=&\delta(p_\eta^1-p_\eta^2)\;\;.
\end{eqnarray}
Thus $\mathbb{I}=\int_{-\infty}^{+\infty}\, \overline{|\eta\ra}\overline{\la \eta|}d\eta$, while from (\ref{x_inner}), we have
\begin{eqnarray}
\overline{\la \eta_1| \eta_2\ra}&=&\overline{\la \eta_1|}\left(\int_0^\infty dx\,|x\ra\la x|\right) \overline{|\eta_2\ra}\;\;,\\
&=&\int_0^\infty dx\, e^{\eta_1/2+\eta_2/2}\delta(x-e^{\eta_1})\delta(x-e^{\eta_2})\;\;,\\
&=& e^{\eta_1/2+\eta_2/2}\delta(e^{\eta_1}-e^{\eta_2})\;\;,\\
&=& \delta(\eta_1-\eta_2)\;\;.
\end{eqnarray}
One can then easily show that the new coordinate kets $\overline{|\eta\ra}$, (where now $\eta\in\mathbb{R}$), are completely analogous to $|x\ra$, on the entire real line, and the dilatonic momentum $\peta$, (which generates a dilation of $x$),  generates displacements of $\eta$:
\begin{equation}
\overline{\la \eta|} \peta |\psi\ra=-i\hbar\frac{d}{d\eta}\overline{\la \eta|}\psi\ra\;\;,
\end{equation}
and now we can make the definition of $\hat{\eta}$, (which we previously had as $\hat{\eta}\sim \ln \hat{x}$), more precise:
\begin{equation}
\hat{\eta}=\int_{-\infty}^{+\infty}d\eta\, \eta \overline{|\eta\ra}\overline{\la \eta|}\;\;.
\end{equation}
Armed with the conjugate exponential operators, $\hat{\eta}$, and $\peta$, where $[\et, \peta]=i\hbar$, and the transition matrix elements $\la \eta |x\ra$, $\la p_\eta | x\ra$, we can take any quantum state represented in the $x$ coordinates and represent it in the dilation coordinates: $\la x_1|\rho| x_2\ra\rightarrow \overline{\la \eta_1 |}\rho\overline{|\eta_2\ra}$. 


\section{Mellin transforms and the Lerch Transcendent}
As we mentioned above, one is typically familiar with the Fourier transform appearing in quantum mechanics representing the unitary change in basis $|x\ra\rightarrow|p\ra,\;\;x,p\in\mathbb{R}$. To the best of our knowledge the Mellin transform has never appeared in the literature playing a similar role. However, as we will see, the above change in basis $|x\ra\rightarrow |p_\eta\ra$, yields a Mellin transform. The Mellin transform and its inverse, of a function $f$ is given by
\begin{eqnarray}
\left\{{\cal M}f\right\}(s)&=&\phi(s)=\int_0^\infty\,x^sf(x)\frac{dx}{x}\;\;,\\
\left\{{\cal M}^{-1}\phi\right\}(x)&=&f(x)=\frac{1}{2\pi i}\int_{c-i\infty}^{c+i\infty}\,x^{-s}\phi(s)dx\;\;.
\end{eqnarray} 
The Mellin transform may also be defined via the Fourier transform via,
\begin{equation}
\left\{{\cal M}f\right\}(s)=\left\{{\cal F}f(e^{-x})\right\}(-is)\;\;.
\end{equation}
From (\ref{p_eigen1}), we see that the representation of a given wavefunction $\la x|\psi\ra$, in the $|p_\eta\ra$, basis is,
\begin{eqnarray}
\la p_\eta |\psi\ra &=& \int_0^\infty dx\, \la p_\eta |x\ra\la x|\psi\ra\;\;,\\
&=&\frac{1}{\sqrt{2\pi}}\int_0^\infty dx\, \psi(x) x^{-1/2-ip_\eta}\;\;,\\
&=&\frac{1}{\sqrt{2\pi}}\left\{{\cal M}\psi\right\}(s=1/2-ip_\eta)\;\;.\label{mellineval}
\end{eqnarray}
We now return to our previous expression for the Lerch transcendent (\ref{mellinlerch}), we see that
\begin{eqnarray}
\Xi(z,s=1/2-ip_\eta,u)&\equiv&\Gamma(s=1/2-ip_\eta)\Phi(z,s=1/2-ip_\eta,u)\\
&=&\int_0^\infty dt\,
\frac{e^{-(u-1)t}}{e^t-z}\,t^{-1/2-ip_\eta}\;\;,\\
&=&\left\{{\cal M} \chi\right\}(z,s=1/2-ip_\eta,u)\;\;,
\end{eqnarray}
where
\begin{equation}
\chi(z,t,u)=\frac{e^{-(u-1)t}}{e^t-z}\;\;.\label{integrand}
\end{equation}
Thus, to simulate the three parameter family of Lerch transcendents $\Xi(z,1/2-ip_\eta,u)$, we take the Mellin transform of $\chi(z,t,u)$, evaluated on the critical line $s=1/2-ip_\eta$. However, by (\ref{mellineval}), this can be viewed as the $|p_\eta\ra$ representation of a wavefunction $\psi$, where
\begin{equation}
\la x|\psi(z,u)\ra=\psi_{z,u}(x)= N(z,u)\frac{e^{-ux}}{1-ze^{-x}}\;\;,\label{wavef}
\end{equation}
with the normalization factor $N(z,u)$, and where $x\in\mathbb{R}^+$. We can now ask about the quantum system where the wavefunction (\ref{wavef}), could appear. More specifically, can we find a potential $V(z,u; x)$, for which (\ref{wavef}) is an eigenstate of the half-line Schroedinger equation,
\begin{equation}
\left[-\frac{\hbar^2}{2}\frac{d^2}{dx^2}+V(x)\right]\psi_{z,u}(x)=E\psi_{z,u}(x)\;\;.
\label{schroe}
\end{equation}
As we have noted in section (\ref{QMinhalfline}), the Hamiltonian/Schroedinger operator is only Hermitian and self-adjoint if one restricts the domain of the wavefunctions to satisfy $\psi_{z,u}^\prime/\psi_{z,u}|_{x=0}=\kappa\in\mathbb{R}$. From (\ref{wavef}), we can find
\begin{equation}
\lim_{x\rightarrow 0^+}\,\frac{\psi_{z,u}(x)^\prime}{\psi_{z,u}(x)}=\kappa=-u+\frac{z}{z-1}\;\;.
\end{equation}
Inserting (\ref{wavef}), into (\ref{schroe}), and letting $R=ze^{-x}/(1-ze^{-x})$, we can  solve for the potential function,
\begin{equation}
V(x)=E+\frac{1}{2}\left[ (u+R)^2+R(R+1)\right] \;\;.\label{eigen1}
\end{equation}
From the form of (\ref{eigen1}), we can add a constant to $E$, without changing the condition that (\ref{wavef}) solves (\ref{schroe}). We add $-(E+u^2/2)$, so that $\lim_{x\rightarrow\infty} V(x)=0$, and thus we can set
 \begin{equation}
\bar{V}(x)=+\frac{1}{2}\left[ (u+R)^2-u^2+R(R+1)\right] \;\;,\label{eigen2}
\end{equation}
where now $E=-u^2/2$, and  $\lim_{x\rightarrow\infty}\bar{V}(x)=0$. We note that 
\begin{equation}
\lim_{x\rightarrow 0^+} \bar{V}(x)=-\frac{1}{2}\frac{uz}{z-1}+\frac{1}{2}\frac{z(1+z)}{(z-1)^2}\;\;,\label{Lerchpot}
\end{equation}
and this becomes singular when $z=1$. This singularity at $z=1$ is to be expected as our analogy breaks down since (\ref{mellinlerch}) holds for $z=1$, only when $\Re(s)>1$, but above we assumed $s=1/2-ip_\eta$. 

A particularly interesting case is when $z=-1,\;u=1$, as the Mellin transform of the wavefunction 
\begin{equation}
\la x|\psi_{\zeta}\ra={\cal N}\frac{1}{1+e^x}\;\;,
\end{equation}
where  the normalisation factor ${\cal N}^{-2}=-1/2+\ln 2$,  gives via (\ref{alterzeta}), the Riemann Zeta function
\begin{equation}
\la p_\eta|\psi_\zeta\ra={\cal N}(1-2^{1/2+ip_\eta})\Gamma(1/2-ip_\eta)\zeta(1/2-ip_\eta)/\sqrt{2\pi}\;\;.\label{RZpsi}
\end{equation}
However the amplitude of this wavefunction, now represented in the Hyperbolic momentum coordinate $p_\eta$, falls off rapidly with $p_\eta$, due to the $\Gamma$-function in (\ref{RZpsi}).
From (\ref{Lerchpot}), we see that $\la x |\psi_\zeta\ra$, is an eigenfunction of the potential
 \begin{equation}
 V_{\zeta}(x)=-\frac{1}{2}\left[1-\frac{e^x}{e^x+1}\tanh \frac{x}{2}\right]\;\;,\;\;x\in\mathbb{R}^+\;\;,\label{eigenpot}
 \end{equation}
 with the boundary condition 
 \begin{equation}
 \lim_{x\rightarrow 0^+}\frac{\psi_\zeta(x)^\prime}{\psi_\zeta(x)}=-\frac{1}{2}\;\;.\label{boundaryc}
 \end{equation}
The prefactors  appearing in (\ref{RZpsi}), possesses no zeros and thus the location of the zeros of the wavefunction $\la p_\eta|\psi_\zeta\ra$, corresponds exactly with those of the Riemann Zeta function. However these prefactors, $(1-2^{1-s})\Gamma(s)$, when $s=1/2-ip_\eta$, damp as $\exp(- p_\eta)$. Thus the resulting wavefunction is mostly peaked around $p_\eta\sim 0$, and the finer details of the zeros of the wavefunction are very difficult to observe for large $p_\eta$. For clarity, since $\la p_\eta|\psi\rangle\sim \zeta(1/2-ip_\eta)\,e^{- p_\eta}$, in order to distinguish the zeros of the wavefunction we plot  in Fig. \ref{fig:fig1}, the magnitude of the unnormalised wavefunction (in a logarithmic scale), rescaled by $p_\eta$. We see perfect correspondence between the wavefunction and the Zeta function zeros. Below we will show other forms for $\psi_\zeta(x)$, which yield much closer approximations to the Zeta function, i.e. where the prefactors vary more slowly with the Hyperbolic momentum $p_\eta$.
 
 \section{Exact quantum simulation of the $\zeta$ function}
 \par From Fig. \ref{fig:fig1}, we see good correspondence between $\psi_\zeta(p_\eta)$, and the Riemann Zeta function on the critical line. However the correspondence is detracted by the exponential damping given by the $\Gamma$-function appearing in (\ref{alterzeta}). It would be preferable if one could find a wavefunction in the $x-$representation $\la x|\psi\ra$, whose representation in the Hyperbolic momentum $\la p_\eta|\psi\ra$, gives a closer match to the Riemann Zeta function than afforded by (\ref{alterzeta}). Surprisingly a wavefunction which yields a perfect match can be found. To describe this wavefunction we make reference to a set of qubit basis states proposed by Gottesman, Kitaev and Preskill (GKP), \cite{gottesman01}, for quantum information processing using continuous variables.  On the entire line, $x\in\mathbb{R}$ these states are defined by
 \begin{eqnarray}
 \overline{\overline{|0\ra}}&=&\sum_{n=-\infty}^{+\infty}\,|x=2\theta n\ra =\sum_{n=-\infty}^{+\infty}\,|p=\pi\theta^{-1}n\ra\;\;,\\
 \overline{\overline{|1\ra}}&=&\sum_{n=-\infty}^{+\infty}\,|x=2\theta n+\theta\ra =\sum_{n=-\infty}^{\infty}\,(-1)^n|p=\pi\theta^{-1}n\ra\;\;,
 \end{eqnarray}
$\theta\in[0,1]$, which are infinite superpositions of infinitely squeezed states (i.e. eigenstates of $\hat{x}$, and $\hat{p}$), which are unnormalised and also possess infinite energy. Such continuous variable states have been proposed  to encode a qubit $|q\ra=a\overline{\overline{|0\ra}}+b\overline{\overline{|1\ra}}$ \cite{gottesman01}, and recently an error analysis \cite{glancy06}, and various physical methods for their synthesis \cite{giovannetti01,pinard05,travaglione02,pirandola04,pirandola06a,pirandola06b}, have appeared in the literature. As we are concerned with quantum mechanics on the half-line we set $\theta=1/2$, and only consider the $\delta$-function sum in $\overline{\overline{|0\ra}}$, over a strictly positive domain,
 \begin{equation}
 \widetilde{|0\ra}=\sum_{n=1}^{+\infty}\, |x=s\ra\;\;.
 \end{equation}
Following from (\ref{p_eigen1}), we can easily find
\begin{equation}
\la p_\eta \widetilde{|0\ra} \sim \sum_{n=1}^\infty\,\frac{1}{n^s}\;\;,
\end{equation}
where $s=1/2-ip_\eta$, which is precisely the Riemann Zeta function evaluated on the critical line. However, just as the state $\widetilde{|0\ra}$, in the $x-$representation is unbounded and possesses infinite energy, the Riemann Zeta function for $\Re(s)<1$ is also formally unbounded and is defined there via analytic continuation. Thus although formally we have a wavefunction whose $p_\eta$-representation is precisely the Zeta function, the unboundedness poses severe difficulties in its physical interpretation. 

\section{Intermediate simulation}
We have seen a wavefunction whose $p_\eta$-transform yields the Riemann Zeta function up to a rapidly damped multiplicative function and another wavefunction whose $p_\eta$-transform yields the Riemann-Zeta function precisely.
We now instead search for an intermediate  wavefunction in the $x-$representation which does not possess any pathologies such as unbounded norm, infinite energy etc  and which gives the Zeta function up to some multiplicative factor which damps far slower than above, i.e. slower than $\Gamma(1/2-ip_\eta)\sim \exp(-p_\eta)$. As we shall see, there are an infinite number of such examples. For the most part however, we will be unable to give simple analytic formulae  for these wavefunctions.

\begin{figure}
\begin{center}
\setlength{\unitlength}{1cm}
\begin{picture}(9,13)
\put(-2,7.3){\includegraphics[width=15.2cm,height=5cm]{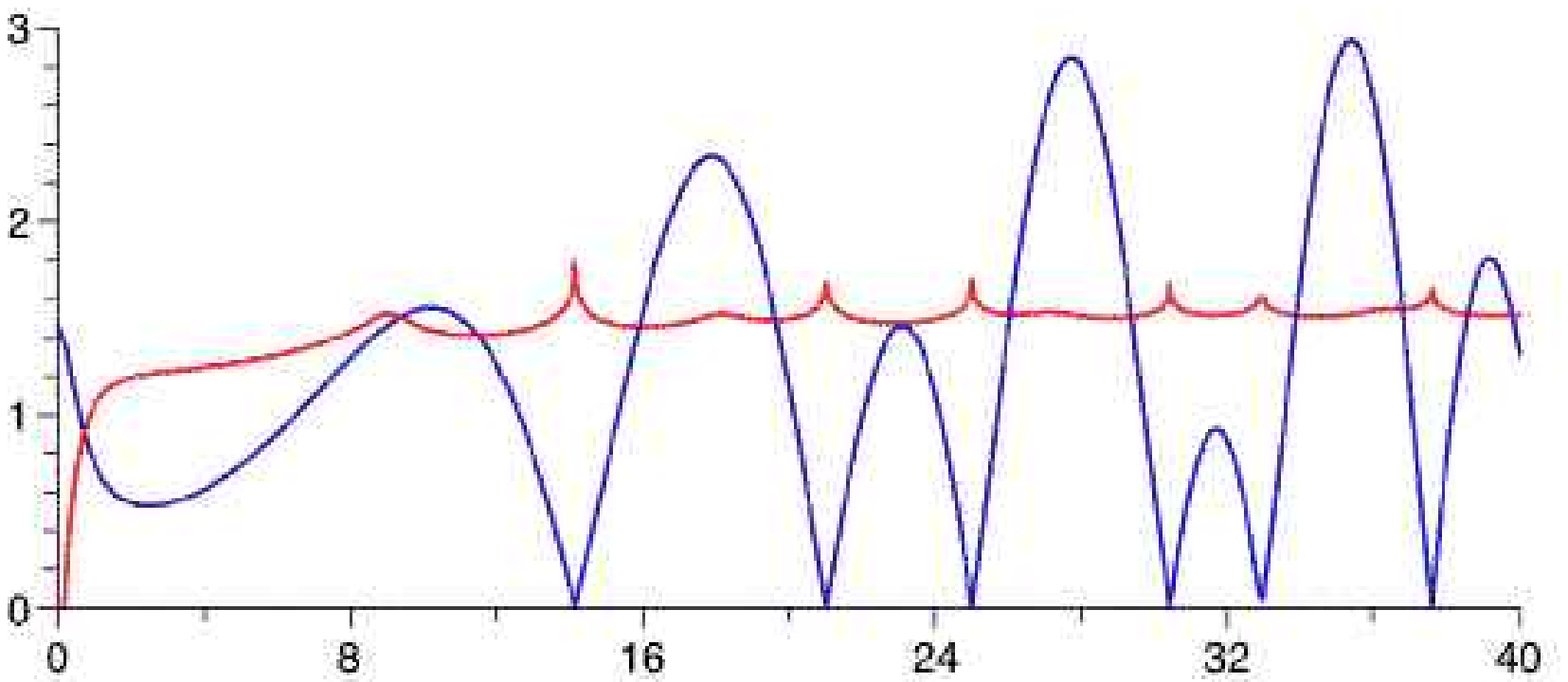}}
\put(5.5,7){\large$-p_\eta$}
\put(.4,10){(B)}
\put(.4,8){(A)}
\put(1,11.5){\Large (I)}
\put(-1,0.5){\includegraphics[width=6cm,height=5cm]{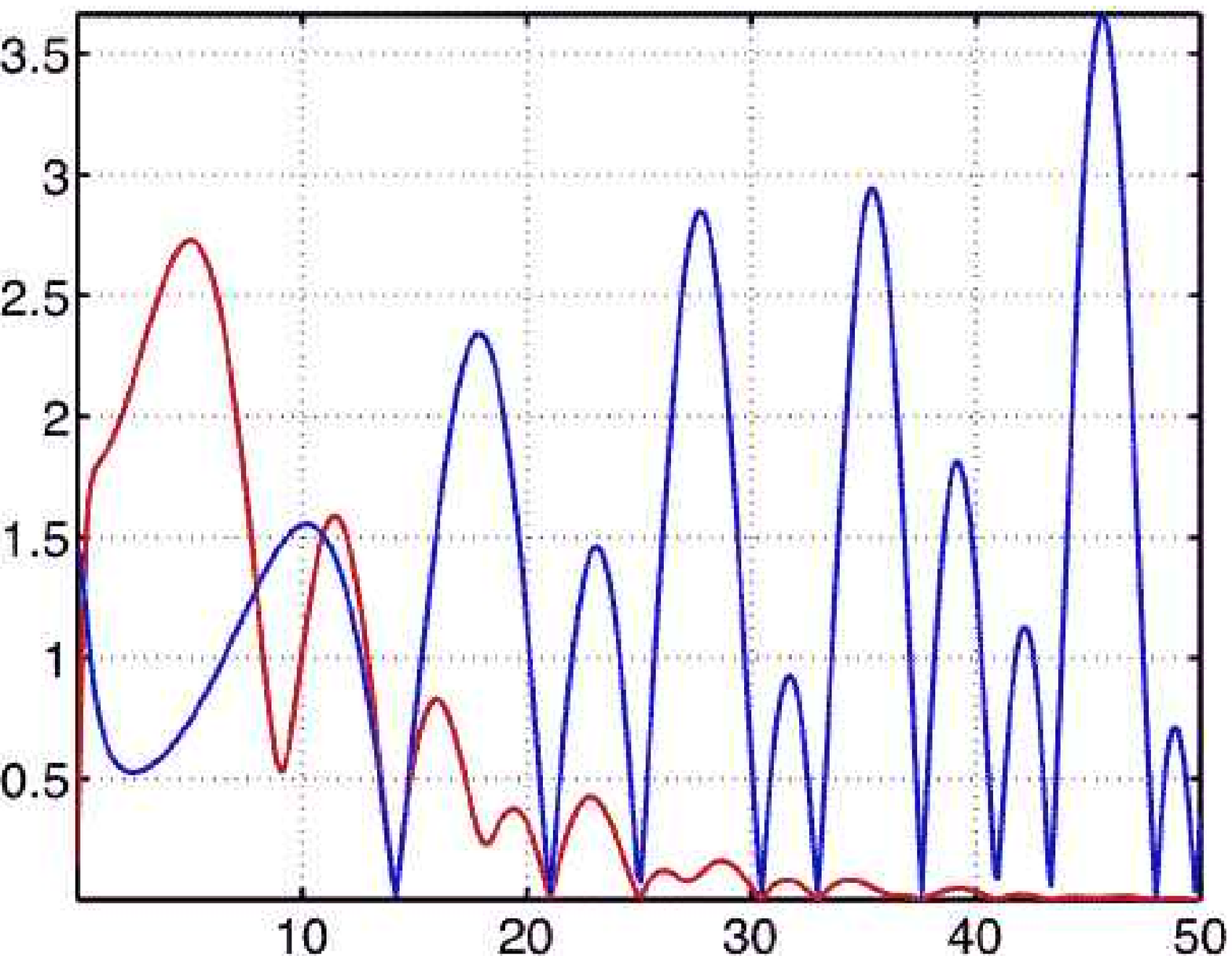}}
\put(6,0.5){\includegraphics[width=6cm,height=5cm]{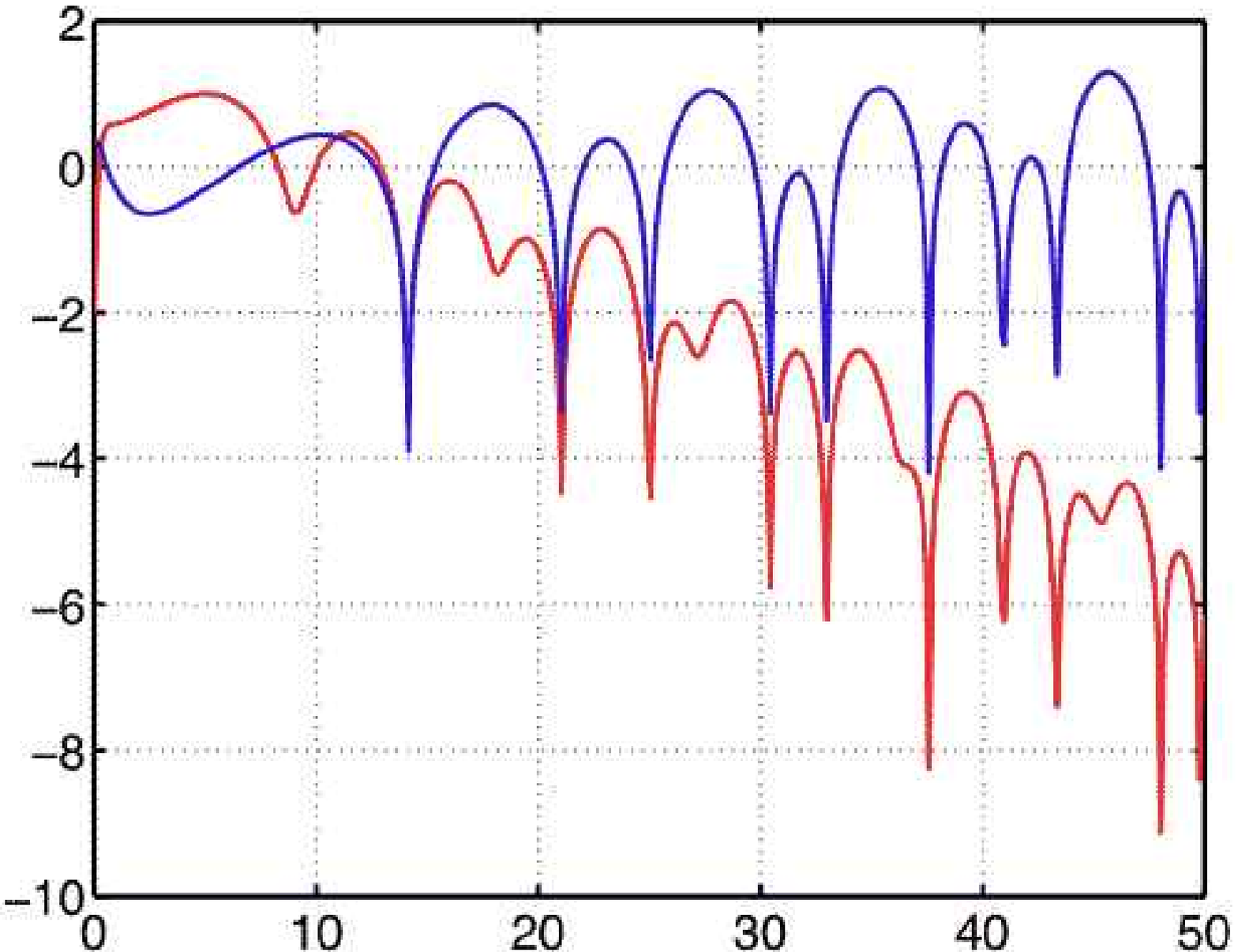}}
\put(.1,4.8){\Large (IIA)}
\put(7,1){\Large (IIB)}
\put(2.,0){\large $-p_\eta$}
\put(9.,0){\large $-p_\eta$}
\end{picture}
\end{center}
\caption{Comparison of  $|\zeta(s=1/2-ip_\eta)|$ (blue), and $|\psi_\zeta(p_\eta)|$ (red) wavefunction in (I) from Eq.(\ref{RZpsi}) and (II) from Eq.(\ref{finalzeta}). In (I) we plot $[\log |\psi_\zeta(p_\eta)|]/(-p_\eta)$ (red), to overcome the exponential supression of (\ref{RZpsi}), generated by the $\Gamma(1/2-ip_\eta)$ function while in (IIA) we plot $|\psi_\zeta(p_\zeta)|$ (red) on a linear vs. linear scale and in (IIB) we repeat on a log vs. linear scale. In (IIA) and (IIB), the details of the zeros of the wavefunction are more clearly visible for a larger range of $p_\eta$.
\label{fig:fig1}}
\end{figure}

\begin{figure}
\begin{center}
\setlength{\unitlength}{1cm}
\begin{picture}(12,12)
\put(-.5,0){\includegraphics[width=7cm,height=6cm]{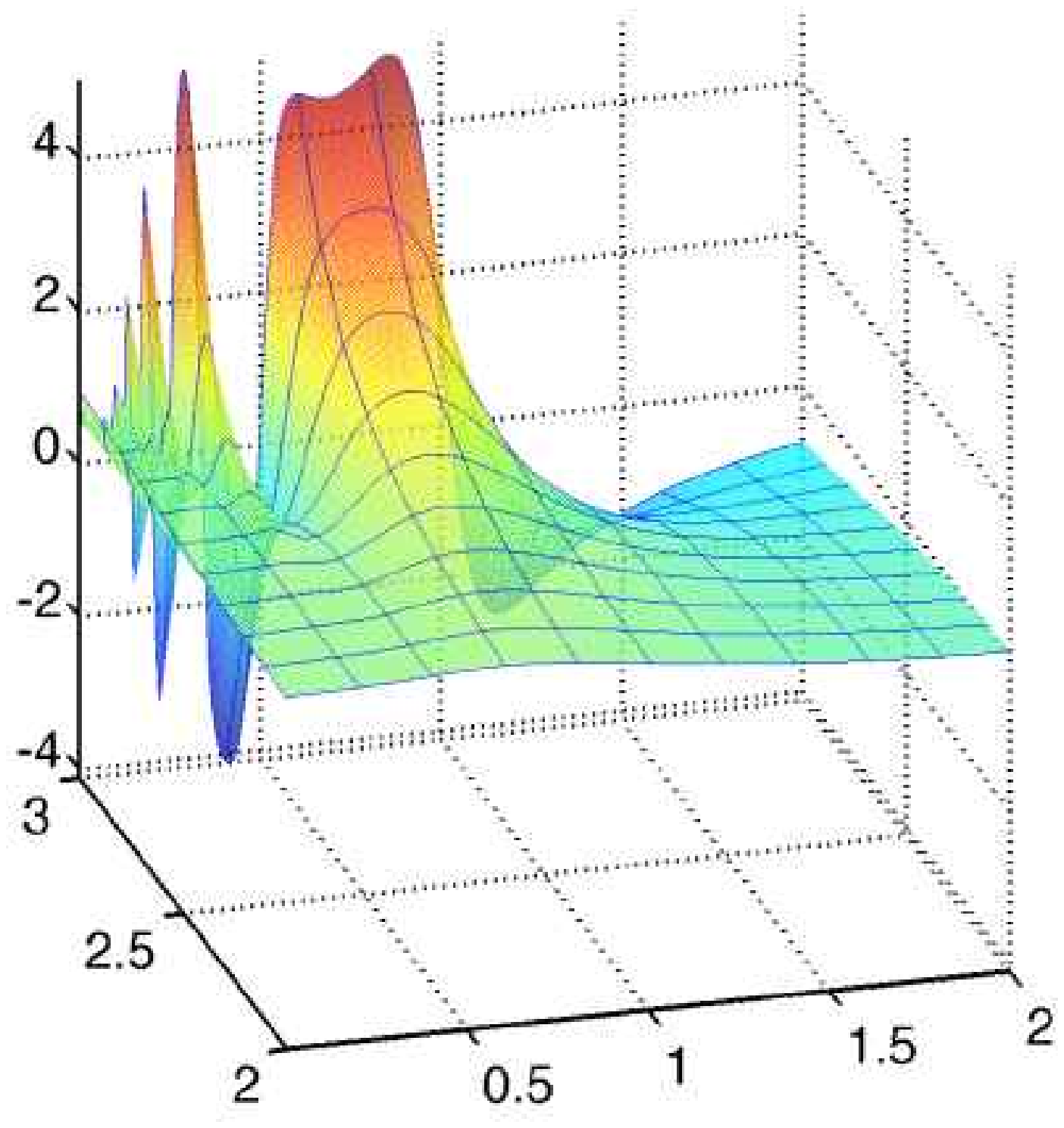}}
\put(4,-.3){\large $x$}
\put(-.9,3.5){\large$\Sigma$}
\put(-.1,.6){\large$\phi$}
\put(7,0){\includegraphics[width=7cm,height=6cm]{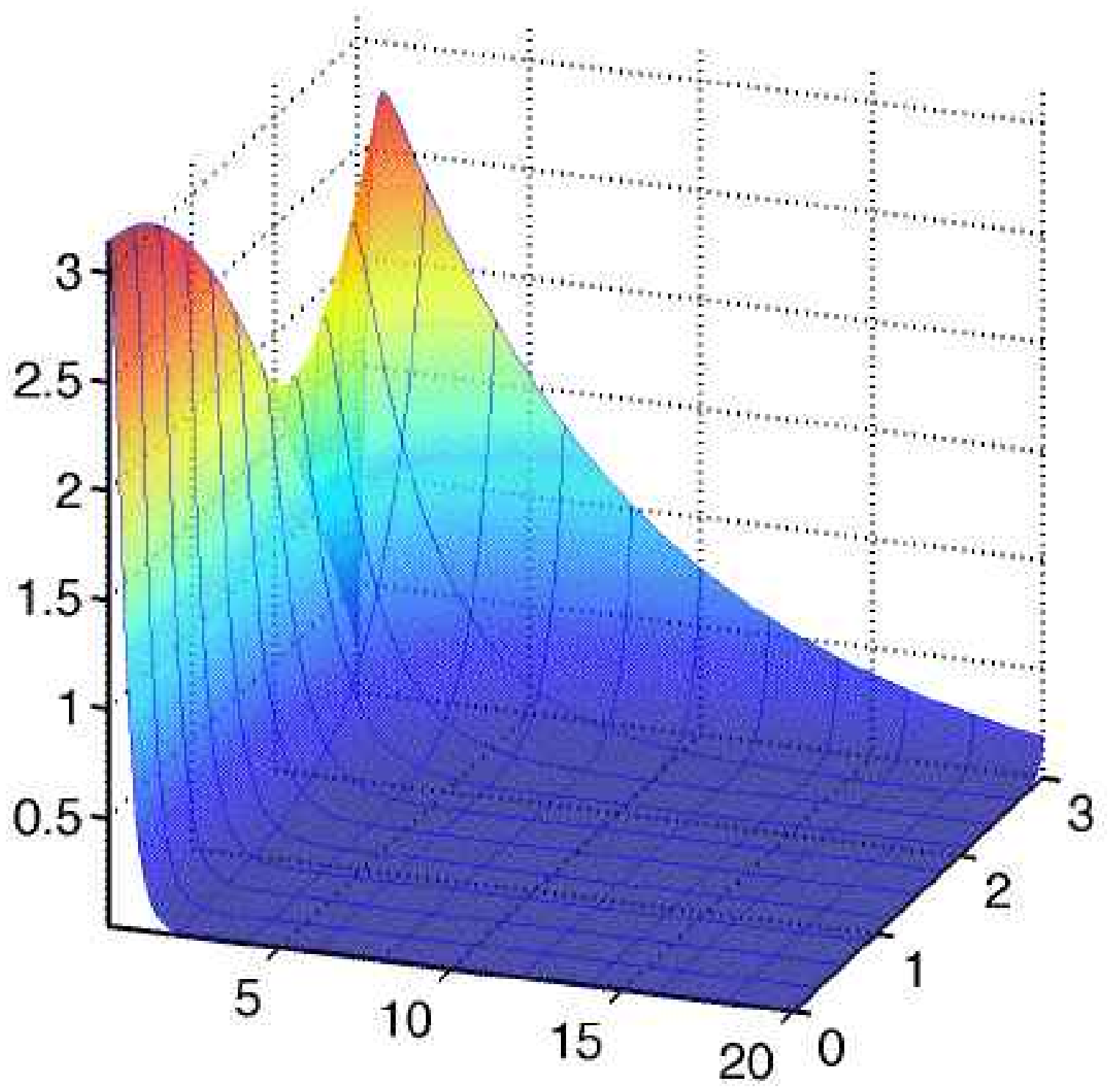}}
\put(13.3,.8){\large $\phi$}
\put(6.8,1.2){\begin{sideways} $|\{{\cal M} f\}(1/2+it)|$\end{sideways}}
\put(9.5,.1){\large$t$}
\put(0,6){\Large(A)}
\put(7.5,6){\Large(B)}
\end{picture}
\end{center}
\caption{Graphs of (A) the sum $\Sigma(x,\phi)=\sum_{n=1}^\infty\,(-1)^n f(nx)$, from Eq. (\ref{Sigma}), and (B) the Mellin transform of $f(x)$, $|\Xi(s,\phi)|$ from Eq. (\ref{melling}), for $s=1/2+it$.
For low values of $\phi\sim 2$, $\Sigma(x,\phi)$ possesses very little structure while for $\phi\sim 3$, this function develops more complex oscillations. For low values of $\phi\sim 0$, the Mellin transform $\Xi$ is highly damped function of $t$, while  for $\phi\sim 3$, the transform  damps much more slowly, thus allowing extended explorations of the  Zeta function on the $-p_\eta$ axis.
 \label{fig:fig2}}
\end{figure}

To find these wavefunctions we return to Eqn. (\ref{alterzeta}),
\begin{equation}
(1-2^{1-s})\zeta(s)\Gamma(s)=\int_{0}^{\infty} 
\frac{1} {e^t+1} t^{s-1} dt\;\;.\label{alterzeta1} 
\end{equation}
This is a special case of a more general Mellin transform formula \cite{sneddon72},
\begin{equation}
\left\{{\cal M}\left(\sum_{n=1}^\infty\,(-1)^{n-1}f(xn)\right)\right\}(s)=(1-2^{1-s})\left\{{\cal M} f\right\}(s) \zeta(s)\;\;,\label{snedsum}
\end{equation}
where $f(x)$ is any function. We see that the particular case of (\ref{alterzeta1}), is obtained when we choose 
\begin{equation}
f(x)=\exp(-x)\;\;,\qquad \left\{{\cal M} e^{-x}\right\}(s)=\Gamma(s)\;\;.
\end{equation} 
We thus seek a function (say), $g$, whose Mellin transform $\{{\cal M}s\}(s)$, does not decay rapidly with $t$, for $s=1/2+it$. An example of such a function is the one parameter family of functions
\begin{equation}
g(x,\phi)=N(\phi)\frac{1+x\cos(\phi)}{1+2x\cos(\phi)+x^2}\;\;,\label{myf1}
\end{equation}
where $N(\phi)$ is a normalisation factor. This function has the Mellin transform \cite{sneddon72},
\begin{eqnarray}
\Xi(s,\phi)&\equiv&\left\{{\cal M} g(\phi)\right\}\\
&=&\frac{N(\phi)\pi}{2\sin (\pi s)}\left[(\cos \phi-i|\sin \phi|)^{s}+(\cos \phi +i|\sin \phi|)^{s}\right]\;\;.\label{melling}
\end{eqnarray}
The sum, 
\begin{equation}
\Sigma(x,\phi)\equiv -\sum_{n=1}^\infty\,(-1)^n f(nx)\;\;,\label{Sigma}
\end{equation}
 cannot be easily expressed in analytic form and must represent a normalisable wavefunction on $x\in\mathbb{R}^+$, and thus we choose the normalisation factor $N(\phi)$, such that $\int_0^\infty|\Sigma|^2dx=1$. We plot the behaviour of (the unnormalised), $\Sigma(x,\phi)$, in Fig. \ref{fig:fig2}. As $\phi\rightarrow\pi$, $\Sigma(x,\phi)$, develops an infinite number of oscillations within the domain $x\in[0,1]$, but in the following we will only consider $\phi\in[0,3]$, and here the function $g(x,\phi)$, and its Mellin transform $\Xi(s,\phi)$, are well behaved. 
 We see that the Mellin transform damps quickly with increasing $t$, for low values of $\phi\in[0,2.5]$, but as $\phi\rightarrow 3$, this function damps more slowly. We expect that by increasing the value of the parameter $\phi\rightarrow \pi$, we can extend the Mellin transform's $\Xi(1/2+it,\phi)$, extent to arbitrary large $t$-values at the expense of using highly oscillatory wave-functions $\Sigma(x,\phi)$. If we now follow this argument and choose a wavefunction in the $x-$domain to be $\la x|\chi\ra=N(\phi=3)\Sigma(x,\phi=3)$, we can move to the $p_\eta-$representation to find
\begin{equation}
\la p_\eta | \chi\ra=N(3)(1-2^{1/2+ip_\eta})\zeta(1/2-ip_\eta)\Xi(1/2-ip_\eta,3)\;\;,\label{finalzeta}
\end{equation}
and we plot $|\la p_\eta|\chi\ra|$, vs. $-p_\eta$, in Fig. \ref{fig:fig1} (II). Compared with Fig. \ref{fig:fig1} (I), we can clearly see that the wavefunction now has greater extent over the $-p_\eta$ coordinate and more details of the zeros/nodes of this wavefunction are apparent.

Thus although we have shown that one can find legitimate normalisable wavefunctions on the $x\in\mathbb{R}^+$, axis whose $p_\eta$-representation gives closer and closer approximations to the Riemann Zeta function, these wavefunctions become more and more pathologic e.g. $\sim \sin(1/x)$.

\section{Wigner Functions in the $(\eta,p_\eta)$ phase space}
\begin{figure}
\begin{center}
\setlength{\unitlength}{1cm}
\begin{picture}(9,9)
\put(1.5,0){\includegraphics[width=8cm,height=8cm]{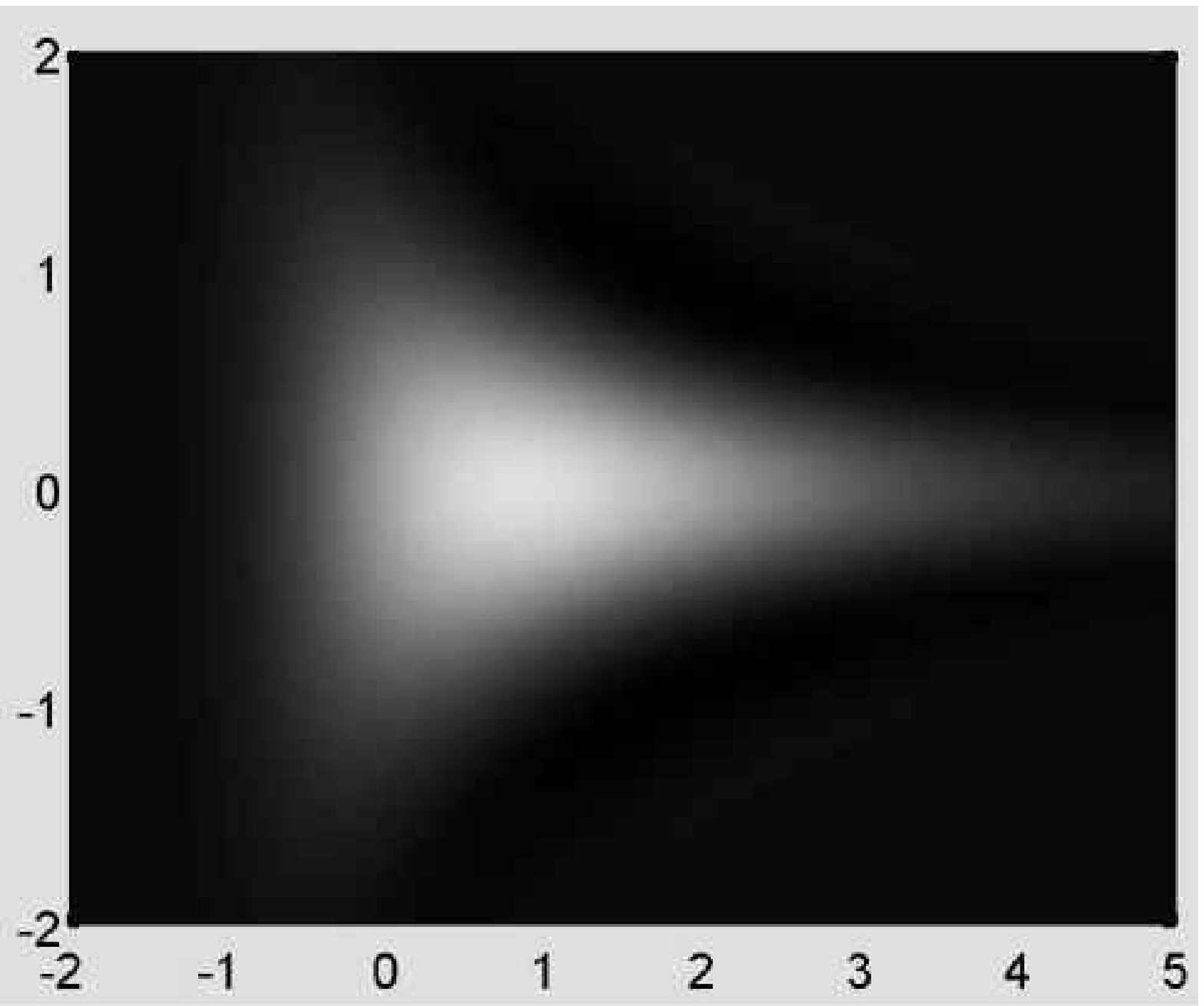}}
\put(5.5,-.25){\large$\eta$}
\put(1,4){\large $p_\eta$}
\end{picture}
\end{center}
\caption{Hyperbolic phase space Wigner function $W(\eta, p_\eta)$, of the Riemannn-Zeta wavefunction in Eq. (\ref{RZpsi}).} 
 \label{fig:fig7}
\end{figure}

\begin{figure}
\begin{center}
\setlength{\unitlength}{1cm}
\begin{picture}(10,10)
\put(-4.25,10){\includegraphics[width=10.25cm,height=19cm,angle=-90]{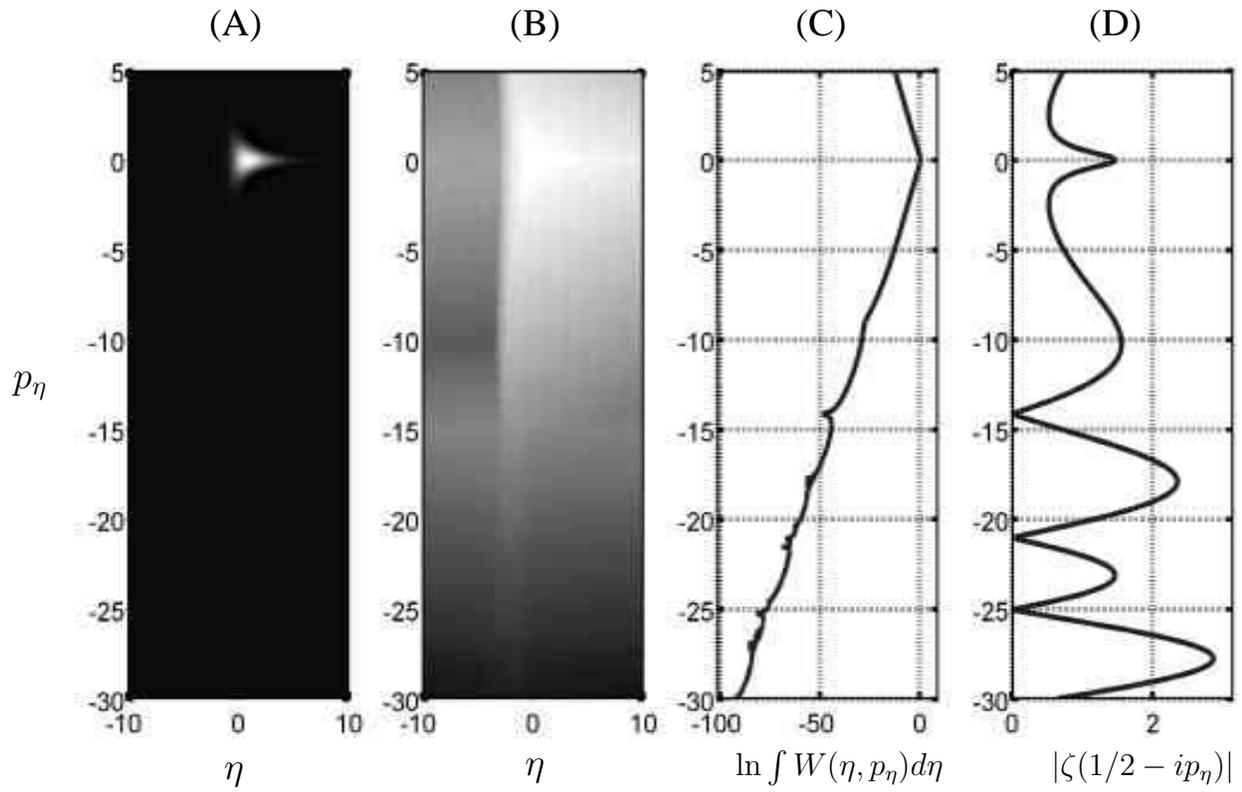}}
\put(-3.3,5){\large $p_\eta$}
\put(-.5,-.2){\large$\eta$}
\put(3.5,-.15){\large$\eta$}
\put(6.3,-.15){$\ln \int W(\eta,p_\eta)d\eta$}
\put(10.5,-.15){$|\zeta(1/2-ip_\eta)|$}
\put(-.7,9.7){\large (A)}
\put(3.3,9.7){\large (B)}
\put(7.1,9.7){\large (C)}
\put(11,9.7){\large (D)}
\end{picture}
\end{center}
\caption{Details of the Hyperbolic phase space Wigner function  $W(\eta, p_\eta)$ of the state (\ref{RZpsi}).  (A) $ W(\eta, p_\eta)$, (B) $\ln|W(\eta,p_\eta)|$, (C) $p_\eta$ vs. $\int W(\eta,p_\eta)d\eta$, and (D) $|\zeta(1/2-ip_\eta)|$. One can see faint traces of the Riemannn-Zeta zeros of the wavefunction within the Wigner function.} 
 \label{fig:fig8}
\end{figure}

\begin{figure}
\begin{center}
\setlength{\unitlength}{1cm}
\begin{picture}(9,9)
\put(1.5,0){\includegraphics[width=8cm,height=8cm]{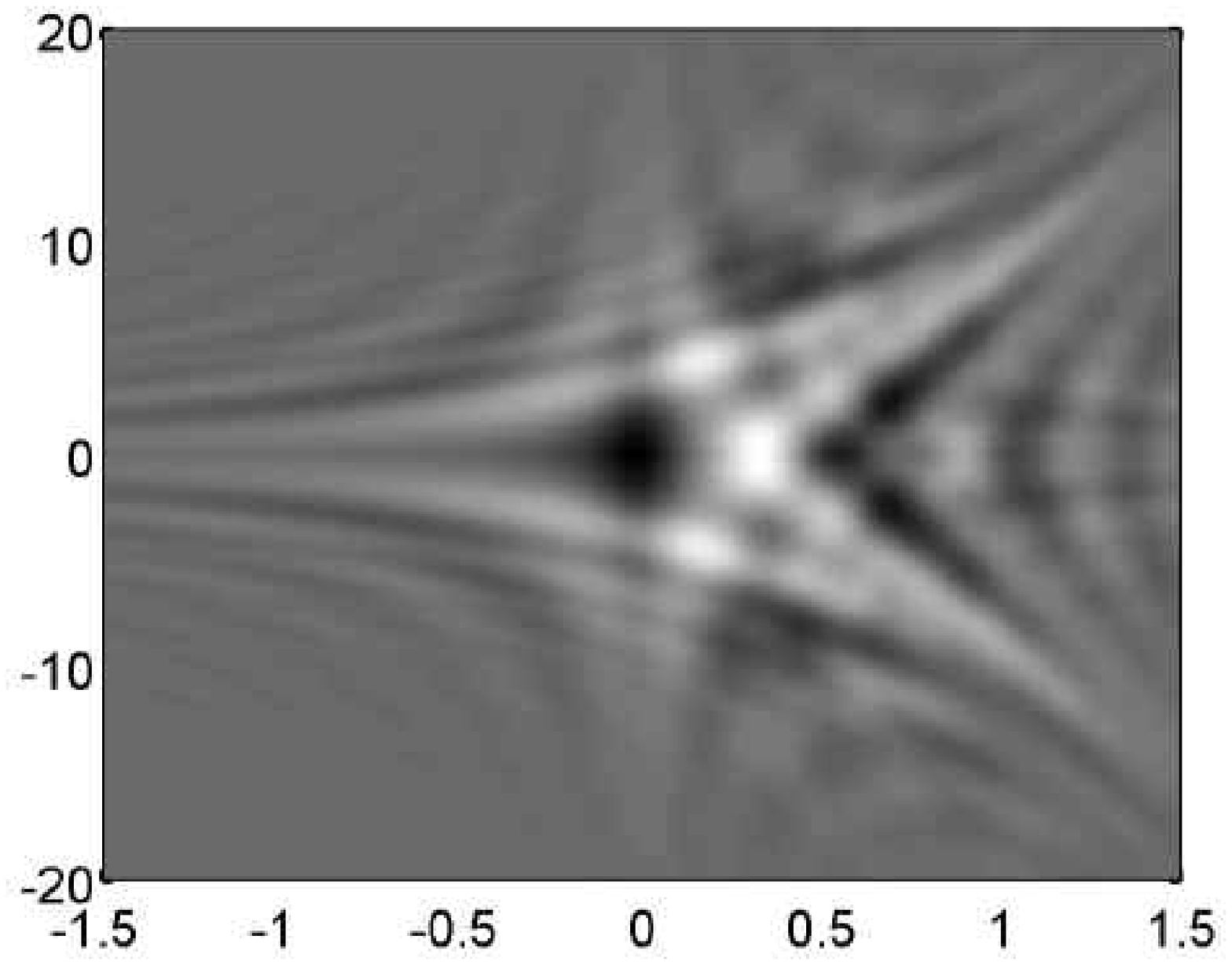}}
\put(5.5,-.25){\large$\eta$}
\put(1,4){\large $p_\eta$}
\end{picture}
\end{center}
\caption{Hyperbolic phase space Wigner function $W(\eta, p_\eta)$, of the Riemannn-Zeta wavefunction in Eq.(\ref{finalzeta}).} 
 \label{fig:fig9}
\end{figure}

\begin{figure}
\begin{center}
\setlength{\unitlength}{1cm}
\begin{picture}(10,10)
\put(-2.3,9.6){\includegraphics[width=9.3cm,height=16cm,angle=-90]{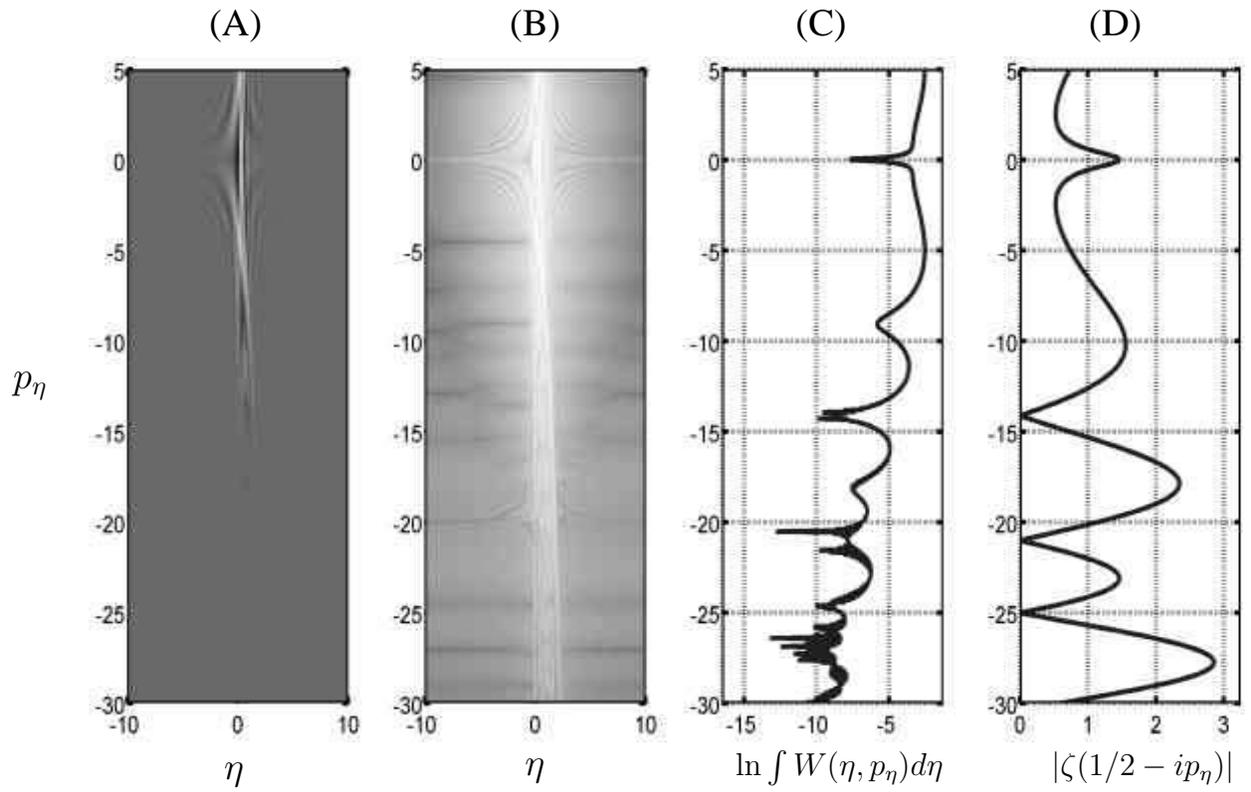}}
\put(-3.3,5){\large $p_\eta$}
\put(-.5,-.2){\large$\eta$}
\put(3.5,-.15){\large$\eta$}
\put(6.3,-.15){$\ln \int W(\eta,p_\eta)d\eta$}
\put(10.5,-.15){$|\zeta(1/2-ip_\eta)|$}
\put(-.7,9.7){\large (A)}
\put(3.3,9.7){\large (B)}
\put(7.1,9.7){\large (C)}
\put(11,9.7){\large (D)}
\end{picture}
\end{center}
\caption{Details of the Hyperbolic phase space Wigner function  $W(\eta, p_\eta)$ of the state (\ref{finalzeta}).  (A) $ W(\eta, p_\eta)$, (B) $\ln|W(\eta,p_\eta)|$, (C) $p_\eta$ vs. $\int W(\eta,p_\eta)d\eta$, and (D) $|\zeta(1/2-ip_\eta)|$. As compared with Fig. \ref{fig:fig8}, the Riemannn-Zeta zeros are more clearly displayed in this wavefunction.} 
 \label{fig:fig10}
\end{figure}

The Wigner quasi-probability distribution function $W(\eta,p_\eta)$, is a standard tool for vizualising the quantum aspects of a quantum state, i.e. when $W$ assumes negative values. The $W$ function however, is typically defined on the phase space $(x,p)$, where $x,p\in \mathbb{R}$. For quantum mechanics on the half-line $x\in\mathbb{R}^+$, we have shown above that one cannot define a self-adjoint momentum operator and thus we cannot construct an associated Wigner function. However in the Hyperbolic representation $(\eta, p_\eta)$, where now both $\eta,p_\eta\in \mathbb{R}$, and $[\hat{\eta},\hat{p}_\eta]=i\hbar$, we can (as in \cite{twamley98}), define a Wigner function to be 
\begin{equation}
W_D(\eta,p_\eta)\equiv \frac{1}{2\pi}\int_{-\infty}^{+\infty}d\eta^\prime\, \overline{\la \eta+\frac{1}{2}\eta^\prime|}
\rho \overline{|\eta- \frac{1}{2}\eta^\prime\ra}\,e^{i\eta^\prime p_\eta}\;\;.
\end{equation}
For the two quantum states  (\ref{RZpsi}), and (\ref{finalzeta}), we have plotted the associated Hyperbolic phase-space Wigner function in Figs. \ref{fig:fig7}-\ref{fig:fig10}. Although zeros of the wavefunction do not normally correspond to zeros of the Wigner function, from Figure \ref{fig:fig8} and Figure \ref{fig:fig10}, some correspondence seems present. Moreover the $p_\eta$-marginal probability, $|\psi_\zeta(p_\eta)|^2=\int W(\eta,p_\eta)d\eta$, exhibits the engineered Riemann-Zeta zeros.

\section{Physical Realisation}
Above we have given a mathematical derivation of a new type of unitary transformation which transforms quantum mechanics on the positive half line to a Hyperbolic representation and in the process, executes a Mellin transform $\psi(x)\rightarrow \left\{{\cal M}\psi\right\}(s=1/2-ip_\eta)/\sqrt{2\pi}$. By choosing $\psi(x)$, carefully (such that it is the eigenstate of a particular potential $V(x)$, (\ref{eigenpot}), subject to the boundary condition (\ref{boundaryc})), the transformed wavefunction $\psi(p_\eta)$, has a nodal structure identical to the Riemann Zeta function on the critical line.  The boundary condition (\ref{boundaryc}), is not typical. However as shown in \cite{pazma89, filop02}, one can closely approximate such a boundary condition by a short-range potential $V_{bndry}(x)$, near the origin. For the boundary condition (\ref{boundaryc}) to hold independent of energy, $V_{bndry}$, limits to a hard-wall infinite potential with infinitesimal structure \cite{filop02} which enforces (\ref{boundaryc}). Possible physical systems which might admit the engineering of such short-range boundary potentials are multi-frequency evanescent-wave mirrors used for trapping and reflecting ultra-cold neutral atoms \cite{cote03}. A more challenging task is the physical realisation of the measurement of Hyperbolic phase $\la p_\eta | \psi\ra$. We propose using an indirect measurement scheme \cite{milburn83,imoto85}, where the system we wish to measure is coupled to a quantum probe via an interaction Hamiltonian $H_I$ such that the Hyperbolic phase of the system drives a linear displacement of the probe's quantum state. Since $[\eta,p_\eta]=1$, we can formulate Hyperbolic creation and annihilation operators $\eta\equiv (\tilde{a}+\tilde{a}^\dagger)/2$, $p_\eta=(\tilde{a}-\tilde{a}^\dagger)/2i$, with $[\tilde{a},\tilde{a}^\dagger]=1$. Taking the probe as a bosonic field described by operators $b,\,b^\dagger,\;[b,b^\dagger]=1$ (i.e. on the full line rather than the half line and these quadrature operators $x_{pb},\,p_{pb}$, are now well defined self-adjoint operators), then the indirect measurement generated by the interaction $H_I^A=\chi(\tilde{a}b^\dagger+\tilde{a}^\dagger b)=2i\chi(x_{pb}p_\eta-p_{pb}\eta)$, will drive displacements of the probe field generated by the system's Hyperbolic quadratures. Unfortunately, although (from (\ref{pdilaton_x})), $p_\eta=[ \x\px+\px\x]/2$, the Hyperbolic position is effectively the logarithm of the position of the system on the half line, $\eta\sim \ln \x$, and since $\ln x \approx \ln a -\sum_{i=1}^\infty\,(-1)^i(x-a)^i/(ia^i)$, $x\le 2a$, the interaction Hamiltonian $H_I^A$ involves coupling the probe to a highly nonlinear function (involving all powers of $\x$), of the system's position. Instead we consider the alternative coupling Hamiltonian 
\begin{eqnarray}
H_I^B&=&2\chi\left[\xpb(\xs^2-\pxs^2)+\ppb(\xs\pxs+\pxs\xs)\right]\nonumber\\
&=&2\chi\left[\xpb(\xs^2-\pxs^2)+2\ppb p_\eta\right]\nonumber\\
&=&\chi(\b^\dagger\a^2+\b\a^{\dagger\,2})\;\;,\label{indirect}
\end{eqnarray} 
where $\xs,\,\pxs,\,\a,\,\a^\dagger$, now describes a bosonic mode defined on the entire real line. We note however that (\ref{indirect}) acts disjointly on the two half-line ``super-selection'' sectors $x\in\mathbb{R}^+$ and $x\in\mathbb{R}^-$.
From this we see that the probe's momentum is coupled to the Hyperbolic momentum in each super-selection sector but the probe's position is no longer coupled to the Hyperbolic position.  Moreover, (\ref{indirect}), is the Hamiltonian for the Degenerate Parametric Oscillator, and has been studied by many \cite{hillery84,hillery94,agarwal97, chaturvedi02, agarwal06}, though mostly in the case where the probe field is treated semiclassically. To get a feeling for how (\ref{indirect}), provides us with an indirect measurement of the Hyperbolic momentum of the system mode we examine the quantum Heisenberg equations of motion and their semiclassical c-number approximations,  
\begin{eqnarray}
\frac{d\a}{dt}&=-2i\chi\b\a^\dagger\;\;,\;\;&\frac{d\alpha}{dt}=-2i\chi\beta\alpha^{*\,2}\;\;,\label{H1}\\
\frac{d\b}{dt}&=-i\chi\a^2\;\;,\;\;&\frac{d\beta}{dt}=-i\chi\alpha^2\;\;,\label{H2}
\end{eqnarray}
given the initial probe state to be the vacuum $|\psi_{pb}\ra=|0\ra$, and the system state only within $\mathbb{R}^+$, i.e. $\la x|\psi_{sys}\ra=0\,\; \forall x\le0$.   Setting $\tau=4\chi t$, quadratures for the probe and system $2\beta=\xb+i\yb,\;2\alpha=\xa+i\ya$,  and introducing the new variables $2u\equiv \xa^2-\ya^2,\;2w\equiv \xa^2+\ya^2,\;v\equiv \xa\ya$, where now $v$ is the semiclassical analogue of the Hyperbolic mometum of the system, equations (\ref{H1},\ref{H2}) can be recast as
\begin{equation}
\frac{d}{d\tau}\left[\begin{array}{c} \xb \\ v\\ w\\ u\\ \yb \end{array}\right]=\left[
\begin{array}{ccccc}
0& 1/2&0&0&0\\
0&0&-\xb& 0&0\\
0&-\xb&0&\yb&0\\
0&0&\yb&0&0\\
0&0&0&-1/2&0
\end{array}\right]\left[\begin{array}{c} \xb \\ v\\ w\\ u\\ \yb \end{array}\right]\;\;.
\label{bigmat}
\end{equation}
From this one can prove that $(\xb^2+\yb^2)+w=2|\beta(t)|^2+|\alpha(t)|^2=K$, is a constant of the motion, while in addition $\ddot{x}_{pb}/\xb=\ddot{p}_{pb}/\yb$. Using this constant of the motion one can arrive at coupled nonlinear dynamics for the probe mode alone,
\begin{equation}
2\ddot{x}_{pb}=-\xb[K-(\xb^2+\yb^2)]\;\;,\;\;2\ddot{y}_{pb}=-\yb[K-(\xb^2+\yb^2)]\;\;,
\end{equation}
with the initial conditions $\xb=\yb=0,2\dot{\xb}=\xa\ya,\;\; 2\dot{p}_{pb}=\xa^2-\ya^2$. This is a conservative system of nonlinear ordinary differential equations which corresponds precisely to the kinematics of a particle moving in two spatial dimensions $(\xb,\yb)$, under the central potential
\begin{equation}
V(\xb, \yb)=\frac{K}{4}r^2-\frac{1}{8}r^4\;\;,
\end{equation}
where $r^2=\xb^2+\yb^2$. The dynamics can be exactly solved via Elliptic functions. One can find a series solution for small times to be
\begin{eqnarray}
\xb(\tau)&=&\frac{1}{2}\xa\ya\tau\left[ 1 -\frac{1}{24}(\xa^2+\ya^2)\tau^2+\cdots\right]\;\;,\\
\yb(\tau)&=&-\frac{1}{4}(\xa^2-\ya^2)\tau\left[1-\frac{1}{24}(\xa^2+\ya^2)\tau^2+\cdots\right]\;\;.
\end{eqnarray}
 Qualitatively the particle moves away from the origin initially with constant velocity (as near the origin the force vanishes), but after a time $t^*\sim \sqrt{\frac{3}{4\w(t=0)}}/\chi$, both $\xb$ and $\yb$ asymptote to fixed values as the particle comes to rest at the maximum of the potential $V$. The initial energy of the system $w(t=0)$, drives the probe up the potential and in the long time limit all of the system's energy is expended into displacing the probe. However the dependence of the probe's displacement on the initial system parameters, $u$ and $v$ is only linear in the small time limit $t\ll t^*$. To summarise, from the semiclassical Heisenberg equations of motion we have good reason to believe that for short interaction times the system's Hyperbolic momentum via, $H_I^B$, induces a linear response of the $\xb$ quadrature. Although Degenerate Parametric Oscillators are studied in optical settings it is also possible to engineer (\ref{indirect}),  between the vibrational modes of an ion in an ion trap \cite{agarwal97}.         

\section{Conclusion}
We have found a new type of unitary transformation for quantum mechanics on the half-line which transforms one into the Hyperbolic representation. The effect of this transformation is to execute a Mellin transform on the wavefunction. Although this work has no immediate impact on the Riemann-Zeta hypothesis, it is interesting that we are able to present bounded quantum wavefunctions whose nodal properties match exactly those of the Riemann Zeta function on the critical line. It is also possible that just as the quantum Fourier transform has found numerous applications in quantum computation, this new quantum Mellin transform might lead to new quantum applications and algorithms.

\bibliography{Zeta_bib4}

\begin{thebibliography}{10}

\bibitem{thompson97}
W.~Thompson, {\it Atlas for computing mathematical functions\/} (John Wiley \&
  Sons, New York, 1997).

\bibitem{bateman53}
H.~Bateman, A.~Erdelyi, {\it Higher Transcendental Functions\/}, vol. Vol. 1
  (McGraw-Hill, New York, 1953).

\bibitem{spanier87}
J.~Spanier, K.~B. Oldham, {\it An Atlas of Functions\/} (Hemisphere, New York,
  1987).

\bibitem{montgomery74}
H.~L. Montgomery, {\it Proc. Int. Cong. Math.\/} (Vancouver, 1974), vol.~I, pp.
  379--381.

\bibitem{odlyzko89}
A.~M. Odlyzko, {\it Supercomputing 89: Supercomputing Structures \&
  Computations, Proc. of 4th Intern. Conf. on Supercomputing\/}, L.~P.
  Kartashev, S.~I. Kartashev, eds. (International Supercomputing Institute,
  1989), pp. 348--352.

\bibitem{rudnick96}
Z.~Rudnick, P.~Sarnak, {\it Duke Mathematical Journal\/} {\bf 81}, 269 (1996).

\bibitem{katz99}
N.~M. Katz, P.~Sarnak, {\it Bulletin of the American Mathematical Society\/}
  {\bf 36}, 1 (1999).

\bibitem{NoteAmerMath03}
{\it Notices Amer. Math. Soc.\/} {\bf 50}, 341 (2003).

\bibitem{twamley98}
J.~Twamley, {\it Journal of Physics A-Mathematical and General\/} {\bf 31},
  4811 (1998).

\bibitem{milburn94}
G.~J. Milburn, W.~Y. Chen, K.~R. Jones, {\it Physical Review A\/} {\bf 50}, 801
  (1994).

\bibitem{carreau90}
M.~Carreau, E.~Farhi, S.~Gutmann, {\it Phys. Rev. D\/} {\bf 42}, 1194 (1990).

\bibitem{daluz94}
M.~G.~E. da~Luz, B.~K. Cheng, {\it Physical Review A\/} {\bf 51}, 1811 (1995).

\bibitem{araujo04}
V.~S. Araujo, F.~A.~B. Coutinho, J.~F. Perez, {\it American Journal of
  Physics\/} {\bf 72}, 203 (2004).

\bibitem{bonneau01}
G.~Bonneau, J.~Faraut, G.~Valent, {\it American Journal of Physics\/} {\bf 69},
  322 (2001).

\bibitem{vonneumann29}
J.~v. Neumann, {\it Math. Annalen\/} {\bf 102}, 49 (1929).

\bibitem{dealfaro76}
V.~d. Alfaro, S.~Fubini, G.~Furlan, {\it Nuovo Cimento A\/} {\bf 34}, 569
  (1976).

\bibitem{Jackiw91}
R.~Jackiw, {\it M. A. B. Beg Memorial Volume\/}, A.~Ali, P.~Hoodbhoy, eds.
  (World Scientific, 1991).

\bibitem{vanwinter98}
C.~van Winter, {\it Journal of Mathematical Physics\/} {\bf 39}, 3600 (1998).

\bibitem{berry99}
M.~V. Berry, J.~P. Keating, {\it Siam Review\/} {\bf 41}, 236 (1999).

\bibitem{berry99a}
M.~V. Berry, J.~P. Keating, {\it Supersymmetry and trace formulae: chaos and
  disorder\/}, V.~Lerner, J.~P. Keating, eds. (Plenum (New York), 1999), pp.
  355--367.

\bibitem{sack58}
R.~A. Sack, {\it Phil. Mag.\/} {\bf 3}, 497 (1958).

\bibitem{gottesman01}
D.~Gottesman, A.~Kitaev, J.~Preskill, {\it Physical Review A\/} {\bf 6401},
  (2001).

\bibitem{glancy06}
S.~Glancy, E.~Knill, {\it Physical Review A\/} {\bf 73},  (2006).

\bibitem{giovannetti01}
V.~Giovannetti, S.~Mancini, P.~Tombesi, {\it Europhysics Letters\/} {\bf 54},
  559 (2001).

\bibitem{pinard05}
M.~Pinard, {\it et~al.\/}, {\it Europhysics Letters\/} {\bf 72}, 747 (2005).

\bibitem{travaglione02}
B.~C. Travaglione, G.~J. Milburn, {\it Physical Review A\/} {\bf 66},  (2002).

\bibitem{pirandola04}
S.~Pirandola, S.~Mancini, D.~Vitali, P.~Tombesi, {\it Europhysics Letters\/}
  {\bf 68}, 323 (2004).

\bibitem{pirandola06a}
S.~Pirandola, S.~Mancini, D.~Vitali, P.~Tombesi, {\it European Physical Journal
  D\/} {\bf 37}, 283 (2006).

\bibitem{pirandola06b}
S.~Pirandola, S.~Mancini, D.~Vitali, P.~Tombesi, {\it Journal of Physics
  B-Atomic Molecular and Optical Physics\/} {\bf 39}, 997 (2006).

\bibitem{sneddon72}
I.~N. Sneddon, {\it The use of integral transforms\/} (McGraw Hill, New York,
  1972).

\bibitem{pazma89}
V.~Pa$\v{z}$ma, P.~Pre$\v{s}$najder, {\it Eur. J. Phys.\/} {\bf 10}, 35 (1989).

\bibitem{filop02}
T.~Fulop, T.~Cheon, I.~Tsutsui, {\it Physical Review A\/} {\bf 66}, 052102
  (2002).

\bibitem{cote03}
R.~Cote, B.~Segev, {\it Physical Review A\/} {\bf 67}, 041604(R) (2003).

\bibitem{milburn83}
G.~J. Milburn, D.~F. Walls, {\it Physical Review A\/} {\bf 28}, 2065 (1983).

\bibitem{imoto85}
N.~Imoto, H.~A. Haus, Y.~Yamamoto, {\it Physical Review A\/} {\bf 32}, 2287
  (1985).

\bibitem{hillery84}
M.~Hillery, M.~S. Zubairy, {\it Physical Review A\/} {\bf 29}, 1275 (1984).

\bibitem{hillery94}
M.~Hillery, D.~Yu, J.~Bergou, {\it Physical Review A\/} {\bf 49}, 1288 (1994).

\bibitem{agarwal97}
G.~S. Agarwal, J.~Banerji, {\it Physical Review A\/} {\bf 55}, R4007 (1997).

\bibitem{chaturvedi02}
S.~Chaturvedi, K.~Dechoum, P.~D. Drummond, {\it Physical Review A\/} {\bf 65},
  033805 (2002).

\bibitem{agarwal06}
G.~S. Agarwal, {\it Physical Review Letters\/} {\bf 97}, 023601 (2006).

\end{thebibliography}

\bibliographystyle{Science}


\begin{scilastnote}
\item This work was supported in part by the European Commission Integrated Project QAP under Contract No. 015848.
\end{scilastnote}


\end{document}